\documentclass[aps,prd,twocolumn,groupedaddress,nofootinbib,floatfix]{revtex4}
\pdfoutput=1

\usepackage[scientific-notation=true]{siunitx}
\usepackage{tikz}
\usetikzlibrary{positioning,shapes,shadows,arrows,decorations.pathreplacing}

\usepackage {natbib}
\usepackage {graphicx}
\usepackage {color}

\usepackage{amsmath, amsfonts, amssymb}
\usepackage{hyperref}

\usepackage[caption=false]{subfig}

\def\commitID{commitID: d84940407420c7b862b6eb8a00b7fcc0940b6759}
\def\commitDATE{ Mon Mar 11 10:10:31 2013 +0100}
\def\commitIDshort{commitID: d849404}
\def\commitSTATUS{CLEAN}

\newcommand{\dcc}{LIGO-P1200185-v2} 

\newcommand{\ltitle}{Fully coherent follow-up of continuous gravitational-wave candidates}
\newcommand{\stitle}{Follow-up of semi-coherent candidates.}

\hypersetup{
  pdftitle={\ltitle},
  pdfauthor={M.\ Shaltev, R.\ Prix},
  pdfkeywords={\$\commitID-\commitSTATUS{} \$}
}

\newcommand{\dopP}{\mathbb{P}}
\newcommand{\dopZ}{\mathbb{Z}}

\newcommand{\mF}{\mathcal{F}}

\newcommand{\mV}{\mathcal{V}}

\newcommand{\Ntemp}{\mathcal{N}}

\newcommand{\Tsft}{T_{\mathrm{SFT}}}
\newcommand{\Nsft}{N_{\mathrm{SFT}}}

\newcommand{\Ndet}{N_\mathrm{det}}

\newcommand{\Tseg}{\Delta T}

\newcommand{\Nseg}{N}

\renewcommand{\vec}[1]{\mathbf{#1}}

\newcommand{\snr}{\rho}
\newcommand{\snrsq}{\snr^{2}}
\newcommand{\SNRSQ}{\mathrm{SNR}^{2}}

\newcommand{\MLE}{\mathrm{MLE}}

\newcommand{\cOSFT}{c^{\mathrm{SFT}}_{0}}

\newcommand{\CC}{C}

\newcommand{\thickness}{\theta}
\renewcommand{\th}{\mathrm{th}}

\newcommand{\maxbbeval}{p}

\newcommand{\Hz}{\mathrm{Hz}}

\newcommand{\mub}{u_{b}}
\newcommand{\mce}{w^{+}}
\newcommand{\mre}{w^{-}}
\newcommand{\minmce}{\mce_{\mathrm{min}}}
\newcommand{\maxmce}{\mce_{\mathrm{max}}}

\newcommand{\nsteps}{n_{\mathrm{steps}}}

\newcommand{\Fth}{\mF_{\mathrm{th}}}
\newcommand{\coFth}{\co{2\mF}_{\mathrm{\th}}^{(\SMC)}}
\newcommand{\coFmax}{\co{2\mF}_{\mathrm{\max}}^{(\SMC)}}
\newcommand{\CGNth}{\co{2\mF}_{\mathrm{\th}}^{(\CGN)}}
\newcommand{\pfA}{p_{\mathrm{fA}}}

\newcommand{\seconds}{\mathrm{s}}
\newcommand{\minutes}{\mathrm{min}}

\newcommand{\days}{\mathrm{d}}
\newcommand{\years}{\mathrm{yr}}

\newcommand{\avg}[1]{\overline{#1}}
\newcommand{\tavg}[1]{\langle#1\rangle}

\newcommand{\detnoise}{S_n}

\newcommand{\Band}{\mathrm{B}}

\newcommand{\mty}{\mu^{*}}

\newcommand{\mm}{\check{\mu}}

\newcommand{\co}[1]{\widetilde{#1}}
\newcommand{\ic}[1]{\widehat{#1}}

\newcommand{\SMC}{S}
\newcommand{\CGN}{G}
\newcommand{\NGO}{\neg\CGN}

\newcommand{\stochastic}{\textbf{stochastic}\ }
\newcommand{\deterministic}{\textbf{deterministic}\ }

\newcommand{\DSNT}{5000 }
\newcommand{\ASNT}{7500 }

\begin{document}

\title[\stitle]{\ltitle}
\author{M.\ Shaltev, R.\ Prix}
\affiliation{Albert-Einstein-Institut, Callinstr.\ 38, 30167 Hannover, Germany}
\date{\commitDATE\\\mbox{\dcc}\\\commitIDshort-\commitSTATUS}


\begin{abstract}
  The search for continuous gravitational waves from unknown isolated sources
  is computationally limited due to the enormous parameter space that  needs to be covered and the
  weakness of the expected signals.  Therefore semi-coherent search
  strategies have been developed and applied in distributed computing
  environments such as Einstein@Home, in order to narrow down the
  parameter space and identify interesting candidates.  However, in
  order to optimally confirm or dismiss a candidate as a possible
  gravitational-wave signal, a fully-coherent follow-up using all the available data is required.

  We present a general method and implementation of a \emph{direct}
  (2-stage) transition to a fully-coherent follow-up on semi-coherent candidates.
  This method is based on a grid-less Mesh Adaptive Direct Search
  (MADS) algorithm using the $\mF$-statistic.
  We demonstrate the detection power and computing cost of this
  follow-up procedure using extensive Monte-Carlo simulations on
  (simulated) semi-coherent candidates from a directed as well as from
  an all-sky search setup.
\end{abstract}

\pacs{XXX} 

\maketitle

\section{Introduction}
\label{sec:intro}
Continuous gravitational waves (CWs) are expected to be emitted from
rapidly spinning non-axisymmetric compact objects, e.g. neutron stars.
 The computational cost of a coherent matched-filtering detection
 statistic, such as the $\mF$-statistic \cite{Jaranowski:1998qm},
is small provided the parameters of the source (i.e.\ sky position
$\alpha,\delta$, frequency $f$, frequency derivatives $\dot{f}$, \ldots) are known.
However, wide parameter-space searches for unknown sources quickly become
computationally prohibitive, due to the large number of points in parameter space
(templates) that need to be searched \cite{Brady:1997ji}.

In order to first reduce the parameter space to smaller, more promising regions, semi-coherent
search techniques have been developed \cite{Brady:1998nj,PhysRevD.72.042004,Krishnan:2004sv,2009PhRvL.103r1102P}
 and are currently being used \cite{2007arXiv0708.3818L,Aasi:2012fw},
 for example in the Einstein@Home distributed computing environment
 \cite{EAH:Misc}.  In a semi-coherent search the
total amount of data $T$ is divided into $\Nseg$  shorter segments of
duration $\Tseg$. The coherent statistics from the individual segments
are combined to a new semi-coherent statistic. At fixed computing cost
these semi-coherent methods are (typically) more sensitive than
fully-coherent searches \cite{PrixShaltev2011}.

Structuring a wide parameter-space search into hierarchical
stages, which increasingly concentrate computational power onto the
more promising regions of parameter space, was first described 
in \cite{Brady:1997ji} and elaborated further 
in \cite{Brady:1998nj},
where a two-stage semi-coherent hierarchical search was considered.
An extended hierarchical scheme with an arbitrary number of
semi-coherent stages and a final  fully-coherent stage was studied
numerically in \cite{PhysRevD.72.042004}, which concluded that three
semi-coherent stages will typically be a good choice.
In \cite{Jaranowski:1999pd} and \cite{Krolak:2004xp}  the use of an optimization procedure
has been considered in the process of estimation of the source parameters, once a candidate
is considered as a detection. In both cases, however, no practical
method or implementation was provided for the systematic
coherent follow-up of semi-coherent candidates.

The aim of the present work is to introduce such a coherent follow-up
search strategy and implementation. This is achieved by exploring the
parameter space around a semi-coherent candidate using a Mesh Adaptive
Direct Search (MADS) algorithm. Using this method, we find that
a fully-coherent follow-up (using all of the available data) of initial
semi-coherent candidates is computationally feasible.

This paper is organized as follows: in Section \ref{sec:cw-searches} we
describe the relevant basic concepts in CW searches, in Section \ref{sec:madsforcgw}
 we propose a search strategy for the systematic follow-up of CW candidates, in Section
\ref{sec:monte-carlo-study} we present a Monte-Carlo study and in Section
\ref{sec:disc} we discuss the results.

\subsubsection*{Notation}
We distinguish a quantity $Q$ when referring to a fully-coherent
stage using a tilde, $\co{Q}$ and when referring to a semi-coherent
stage using an overhat, $\ic{Q}$. Averaging over segments is denoted by an overbar, $\avg{Q}$.

\section{Continuous Gravitational Waves}
\label{sec:cw-searches}

Continuous gravitational-wave signals are quasi-monochromatic and sinusoidal in
the source frame  and undergo phase- and amplitude-modulation due to the
diurnal and orbital motion of the detectors. The phase evolution of the signal
at a detector can be approximated as \cite{Jaranowski:1998qm}
\begin{eqnarray}
\label{eq:1}
 \Phi(t)&\approx&\Phi_{0}
+ 2\pi\sum_{k=0}^{s}\frac{f^{(k)}(t_{0})(t-t_{0})^{k+1}}{(k+1)!}\\\nonumber
&+&2\pi\frac{\vec{r}(t)}{c}\vec{n}\sum_{k=0}^{s}\frac{f^{(k)}(t_{0})(t-t_{0})^{k}}{
k!}\ ,
\end{eqnarray}
where $\Phi_{0}$ is the initial phase, $f^{(k)}\equiv\frac{d^{k}f}{dt^{k}}$ are the
derivatives of the signal frequency $f$ at the solar system barycenter (SSB) at
reference time $t_{0}$, $c$ is the speed of
light, $\vec{r}(t)$ is the vector pointing from the SSB to the detector and
$\vec{n}$ is the unit vector pointing from the SSB to the
gravitational-wave source.

\subsection{Detection statistic}
Following \cite{Jaranowski:1998qm,cutler05:_gen_fstat} the
gravitational-wave response of a detector can be expressed as a sum over four
(detector-independent) amplitude parameters multiplying four
(detector-dependent) basis waveforms. The amplitude parameters can be
analytically maximized over and the resulting detection statistic,
known as the $\mF$-statistic, is therefore a function only of the
template ``phase parameters''
$\lambda \equiv \{\alpha,\delta,f,\dot{f},...\}$, where $\alpha$
(right ascension) and $\delta$ (declination) denote the sky position
of the source.

In the presence of a signal the fully-coherent detection statistic $2\mF$ follows a
non-central $\chi^2$-distribution with 4 degrees of freedom and a non-centrality
parameter given by the squared signal to noise ratio (SNR), $\rho^{2}$. The
expectation value is therefore
\begin{equation}
\label{eq:11}
 E[2\mF] = 4 + \snr^{2}\ ,
\end{equation}
with variance
\begin{equation}
\label{eq:12}
 \sigma^{2}[2\mF] = 2(4+2\snr^{2})\ .
\end{equation}

On the other hand, in the semi-coherent approach we divide the available data into
$\Nseg$ segments of duration $\Tseg$ and combine the individual
coherent statistics of the segments to compute a semi-coherent statistic, namely
\begin{equation}
 \label{eq:13}
  \avg{2\mF}(\lambda)=\frac{1}{N}\sum_{k=1}^{\Nseg}2\mF_{k}(\lambda)\ ,
\end{equation}
where $2\mF_{k}$ is the coherent $\mF$-statistic in segment $k$.
The quantity $\Nseg\,\avg{2\mF}$ follows a non-central $\chi^{2}$-distribution with
$4\Nseg$ degrees of freedom, thus the expectation value  of $\avg{2\mF}$ is
\begin{equation}
 \label{eq:14}
 E[\avg{2\mF}] = 4 + \avg{\snr^{2}}\ ,
\end{equation}
with variance
\begin{equation}
 \label{eq:15}
 \sigma^{2}[\avg{2\mF}] = \frac{2}{N}(4+2\avg{\snr^{2}})\ ,
\end{equation}
where $\avg{\snr^{2}}$  is the average SNR$^2$ over all segments, i.e.
\begin{equation}
 \label{eq:16}
 \avg{\snr^{2}}=\frac{1}{N}\sum_{k=1}^{N}\snr^{2}_{k}\ ,
\end{equation}
and $\snr^{2}_{k}$ denotes the $\SNRSQ$ in segment $k$.

\subsection{Mismatch and Fisher matrix}
A search for sources with unknown signal parameters implies a loss of detection
power compared to the perfectly-matched case.
To quantify this we use the notion of mismatch $\mu$, as first introduced in
\cite{owen96:_search_templates,bala96:_gravit_binaries_metric}. This
is defined as the fractional loss of expected $\SNRSQ$  at some parameter space
point $\lambda$ compared to the expectation $\snr^{2}(\lambda_{s})$ at the signal
 location $\lambda_{s}$, namely
\begin{equation}
\label{eq:17}
 \mu\equiv\frac{\snr^{2}(\lambda_{s})-\snr^{2}(\lambda)}{\snr^{2}(\lambda_{s})}\ ,
\end{equation}
such that $\mu\in[0,1]$. Taylor expansion in small offsets
$\Delta\lambda = \lambda - \lambda_{s}$ around the signal location yields
\begin{equation}
\label{eq:18}
 \mu\equiv g_{ij}(\lambda_{s})\,\Delta\lambda^{i}\Delta\lambda^{j} + \mathcal{O}(\Delta\lambda^{3})\ ,
\end{equation}
where implicit summation over repeated parameter-space indices
$i,j$ applies, and the symmetric positive-definite matrix
$g_{ij}$ is commonly referred to as the parameter-space \emph{metric}.

Neglecting higher order terms, one often uses the ``metric mismatch
approximation'', namely
\begin{equation}
 \label{eq:18b}
 \mty\equiv g_{ij}(\lambda_{s})\,\Delta\lambda^{i}\Delta\lambda^{j}\ ,
\end{equation}
as a distance measure, with a range $\mty\in[0,\infty)$.
This metric mismatch $\mty$ plays an important role in grid-based
searches, where one typically constructs template banks in such way
that the mismatch of any putative signal and the ``closest'' template
is bounded by a \emph{maximal} mismatch $m$, i.e.\
\begin{equation}
 \label{eq:18c}
  \mty \le m\ ,
\end{equation}
everywhere in the template bank.

In the presence of noise, $\mu$ as defined in Eq.~\eqref{eq:17} is not directly accessible, and we
therefore introduce a related quantity, namely the  fractional loss of
\emph{measured} $\SNRSQ$, namely
\begin{equation}
\label{eq:26}
 \mm\equiv\frac{2\mF(\lambda_{s})-2\mF(\lambda)}{2\mF(\lambda_{s})-4}\ .
\end{equation}
Note that $\mm\le1$, but contrary to (\ref{eq:17}) it can also be
(slightly) negative, as we can have $2\mF(\lambda_{s})<2\mF(\lambda)$ due to noise.

For semi-coherent searches the metric is found \cite{Brady:1998nj} as the average of
the fully-coherent metrics over all the segments, namely
\begin{equation}
\label{eq:24}
 \ic{g}_{ij}(\lambda)=\frac{1}{\Nseg}\sum_{k=1}^{\Nseg}g_{ij,k}(\lambda)\ ,
\end{equation}
where $\co{g}_{ij,k}$ is the coherent metric (\ref{eq:18}) in segment $k$.

A standard tool for parameter estimation is
provided by the Fisher information matrix,  which characterizes the statistical
uncertainty of the maximum-likelihood estimators (MLE) $\lambda^{i}_{\MLE}$ for
the signal parameters $\lambda^{i}_{s}$. This can be formulated
\cite{PhysRevD.77.042001,prix06:_searc,SMKVOL1} as the well-known
Cramer-R\'{a}o lower bound on the variance of an unbiased MLE (i.e.\
$E[\lambda^{i}_{\MLE}]=\lambda^{i}_{s}$), namely
\begin{equation}
\label{eq:25a}
\sigma^{2}[\lambda^{i}_{\MLE}] \ge \left\{\Gamma^{-1}\right\}^{ii}\,,
\end{equation}
where the matrix $\left\{\Gamma^{-1}\right\}^{ij}$ denotes the inverse of the Fisher matrix $\Gamma_{ij}$,
which is closely related (e.g.\ \cite{prix06:_searc}) to the metric $g_{ij}$, namely
\begin{equation}
\label{eq:25}
 \Gamma_{ij}=\snrsq \,g_{ij}\ .
\end{equation}
A semi-coherent search over $\Nseg$ segments can be considered as
$\Nseg$ different measurements, thus the semi-coherent Fisher matrix
yields \cite{Coe:2009xf}
\begin{equation}
\label{eq:25+1}
 \ic{\Gamma} = \sum_{k=1}^{\Nseg}\Gamma_{ij,k}\ .
\end{equation}
Assuming constant $\SNRSQ$ for the different segments we can rewrite
\eqref{eq:25+1} in terms of the semi-coherent metric \eqref{eq:24}, namely
\begin{equation}
 \label{eq:25+2}
\ic{\Gamma} = \Nseg\,\avg{\snr^{2}}\,\,\ic{g}_{ij}
\end{equation}
and thus
\begin{equation}
 \label{eq:25b}
\left\{ \ic{\Gamma}^{-1}\right\}^{ij}=\frac{\ic{g}^{ij}}{\Nseg\avg{\snrsq}}\ ,
\end{equation}
where $\ic{g}^{ij}$ is the inverse matrix of $\ic{g}_{ij}$.

\subsection{Computing cost}
The computing cost $\CC$ of a fully-coherent (or an ideal semi-coherent \cite{PrixShaltev2011}) search
is primarily due to the computation of the $\mF$-statistic over all
the templates. For a search over
$\Ntemp$ templates using $\Nseg$ segments of data from $\Ndet$
detectors \cite{PrixShaltev2011}, the computing cost $\CC$ is
\begin{equation}
\label{eq:27}
\CC=\Nseg\Ntemp\Ndet\, c_{1}\ ,
\end{equation}
where $c_{1}$ is the implementation-dependent computing cost for
a single template, segment and detector. 
A method of $\mF$-statistic computation based on short Fourier transforms
(SFTs) \cite{williams99:_effic_match_filter_algor_detec} of length $\Tsft$ is
currently widely used in CW searches and will be considered
in the present work. The cost per template in this case is proportional to the segment duration, namely
\begin{equation}
\label{eq:28}
 c^{\mathrm{SFT}}_{1} = \cOSFT\frac{\Tseg}{\Tsft}\ ,
\end{equation}
where $\cOSFT$ is implementation- and hardware-dependent fundamental computing cost per SFT. 
Using the total number of SFTs
\begin{equation}
\label{eq:29}
 \Nsft = \Nseg\Ndet\frac{\Tseg}{\Tsft}\ ,
\end{equation}
we can write the total computing cost (\ref{eq:27}) of the SFT-method as
\begin{equation}
\label{eq:30}
 \CC = \Ntemp\Nsft\ \cOSFT\ .
\end{equation}

In grid-based searches the number of templates required to cover the search
parameter space $\dopP$ is given by the general expression
\cite{2007arXiv0707.0428P,2009PhRvD..79j4017M}
\begin{equation}
\label{eq:34}
 \Ntemp\equiv\thickness_{n}m^{-n/2}\int_{\dopP} d^{n}\lambda\,\sqrt{\det g}\ ,
\end{equation}
where $\thickness$ is the normalized lattice thickness, $n$ is the
number of search dimensions, $m$ is the maximal template-bank mismatch
(\ref{eq:18c}) and $\det g$ is the determinant of the parameter space
metric (\ref{eq:18}).
The normalized thickness is a constant depending on the grid
structure, e.g. for a hyper-cubic lattice
$\thickness_{\dopZ_{n}}=n^{n/2}2^{-n}$.
The metric $g_{ij}$ depends strongly on the duration $\Tseg$ and the
number of segments $\Nseg$, in such a way that longer observation times
typically require a (vastly) increased number of templates \cite{Brady:1997ji}.

\section{Coherent follow-up of semi-coherent candidates}
\label{sec:madsforcgw}

\subsection{Basic two-stage search strategy}

Here we introduce a simple two-stage strategy for following-up candidates from
semi-coherent searches. In the first stage, called
\textit{refinement}, we employ a finer search using the semi-coherent
statistic $\avg{2\mF}$ to improve the initial maximum-likelihood estimator.
In the second stage, called \textit{zoom}, we apply the fully-coherent
statistic $\co{2\mF}$ using all the data $T$, in order to test
whether the candidate is inconsistent with Gaussian noise and if it
further agrees with the signal model.

\begin{figure}
\begin{tikzpicture}[scale=0.75]


\foreach \x in {0,1,2,3,4} {
            \foreach \y in {0,2,4} {
                 \draw  (\x,\y) circle(2.0pt) [fill=black] {};
       \draw(\x,\y) ellipse [x radius=45pt, y radius=13.0pt, rotate=50];
}
}
\draw (0,4.7) -- (-1.2,5.75) node [above]{semi-coherent isomismatch ellipse};
\draw (4,4) -- (4,5.75) node [above]{grid point};

\draw[dashed,thick] (1.53,2.75) ellipse [x radius=19pt, y radius=7pt, rotate=50];
\draw (1.4,2.45) -- (-1.3,-2.5) node [below]{semi-coherent Fisher ellipse};

\draw  (1.3,2.4) node [] {\Large$\star$};
\draw (1.3,2.4) -- (-3,2.5) node [above]{\large$\lambda_{s}\approx\tilde{\lambda}_{\mathrm{MLE}}$};

\draw  (1,2) circle(3.8pt) [fill opacity=0.7] {};
\draw (1,2) -- (5.5,1.5) node [right]{ \large$\lambda_{c}$};

\draw  (1.55,2.75) node [] {$+$};
\draw (1.55,2.75) -- (5.5,3.5) node [right]{ \large$\hat{\lambda}_{\mathrm{MLE}}$};

%
%
\draw [->,very thick,black] (-1.5,-1.5) -- (5,-1.5) ;
\draw (4.7,-2) node {$\dot{f}$};

\draw[->,very thick,black] (-1.5,-1.5) -- (-1.5,5);
\draw -- (-2,4.6) node {$f$};

\end{tikzpicture}
\caption{2-D search grid in $\{f,\dot{f}\}$ space. The black dots are the search templates, placed such
that the loss of SNR on any putative signal $\lambda_s$ will be
bounded by a maximal mismatch $m$, which
defines the semi-coherent iso-mismatch ellipses. The semi-coherent
Fisher ellipse centered on the MLE $\ic{\lambda}_{\MLE}$ is used to
constrain the zoom parameter space. The aim of the zoom stage is to
find $\co{\lambda}_{\MLE}$}.
\label{fig:searchgrid}
\end{figure}
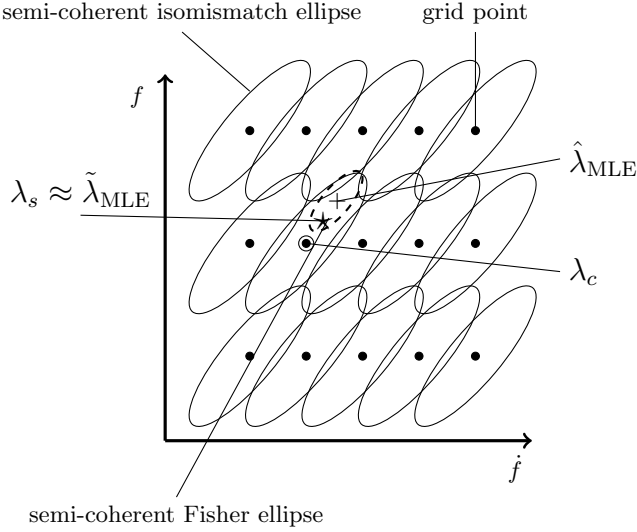

The motivation for this 2-stage approach can be seen from an example
2-D search grid shown in Fig.~\ref{fig:searchgrid}.
The search templates are generally placed such that a putative signal
$\lambda_{s}$ will be recovered with a loss of SNR bounded by a
maximal mismatch $m$, as given in Eq.~(\ref{eq:18c}), namely
\begin{equation}
\label{eq:36}
 g_{ij}\Delta\lambda^{i}\Delta\lambda^{j} \le m\ ,
\end{equation}
where equality defines an ($n$-dimensional) iso-mismatch ellipse.
The initial semi-coherent search will yield ``candidates''
$\ic{\lambda}_{c}$ for which the statistic $2\avg{\mF}$ exceeds a
certain threshold and is higher than neighboring templates.

The initial \emph{refinement} stage of our follow-up strategy therefore consists in
finding the (nearby) parameter-space point $\ic{\lambda}_{\MLE}$ of
the actual (local) maximum in the statistic $2\avg{\mF}(\ic{\lambda})$
(which is smooth function of $\ic{\lambda}$), referred to as the
maximum-likelihood estimator (MLE).
This can be achieved simply by a denser placement of templates using
the original statistic, i.e.\ by keeping the search setup unchanged in
terms of the number and length of segments.

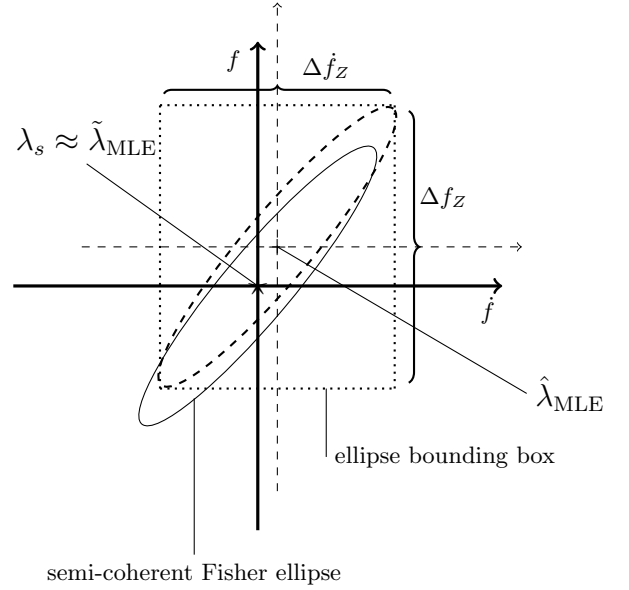
\begin{figure}
\begin{tikzpicture}[scale=0.65]


\draw(0,0) ellipse [x radius=105pt, y radius=20.0pt, rotate=50];

\draw[dashed,thick] (0.4,0.8) ellipse [x radius=105pt, y radius=20pt, rotate=50];
\draw (-1.3,-2.3) -- (-1.3,-5.5) node [below]{semi-coherent Fisher ellipse};

\draw  (0,0) node [] {\Large$\star$};
\draw (0,0) -- (-3.5,2.5) node [above]{\large$\lambda_{s}\approx\tilde{\lambda}_{\mathrm{MLE}}$};

\draw  (0.4,0.8) node [] {$+$};
\draw (0.4,0.8) -- (5.5,-2.2) node [right]{ \large$\hat{\lambda}_{\mathrm{MLE}}$};

%
%
\draw [->,black,very thick] (-5,0) -- (5,0) ; \draw (4.7,-0.5) node {$\dot{f}$};

\draw[->,black,very thick] (0,-5) -- (0,5); \draw -- (-0.5,4.6) node {$f$};

%
%

\draw [->,black,dashed] (-3.6,0.8) -- (5.4,0.8) ;
\draw[->,black,dashed] (0.4,-4.2) -- (0.4,5.8);

%
%


\node (dfdlabel) at (1.4,4.5) {\ \ \ \ \ $\Delta\dot{f}_{Z}\ \ \ \ \ $};
\node [minimum width=3.0cm](dfd) at (0.4,4.3) {};
\draw[decorate,decoration={brace,mirror,raise=-3.5pt,amplitude=3pt}, thick]
     (dfd.south east)--(dfd.south west) ;

\node (dflabel) at (3.8,1.8) {$\Delta f_{Z}$};
\node [minimum height=3.6cm] (df) at (3.4,0.8) {};
\draw[decorate,decoration={brace,mirror,raise=-2.5pt,amplitude=4pt}, thick]
     (df.south west)--(df.north west) ;

%
%

\draw [-,black,dotted,thick] (-2.0,3.7) -- (2.8,3.7);
\draw [-,black,dotted,thick] (2.8,-2.1) -- (2.8,3.7);
\draw [-,black,dotted,thick] (-2.0,-2.1) -- (2.8,-2.1);
\draw [-,black,dotted,thick] (-2.0,-2.1) -- (-2.0,3.7);
\draw (1.4,-2.1) -- (1.4,-3.5) node [right]{ ellipse bounding box};
\end{tikzpicture}
\caption{2-D example: Fisher ellipse \eqref{eq:37} defining the zoom
  search space, centered on the semi-coherent MLE $\ic{\lambda}_{\MLE}$.
  The extents $\{\Delta f,\Delta \dot{f}\}$ of the bounding box are
  given by Eq.~\eqref{eq:5}.}
\label{fig:searchbox}
\end{figure}

In the \emph{zoom} stage we fully-coherently search the Fisher ellipse
centered on the semi-coherent MLE $\ic{\lambda}_{\MLE}$. This defines the
parameter-space region that should contain the signal location
$\lambda_{s}$ with confidence corresponding to $n_{\Band}$ standard
deviations, i.e.\
\begin{equation}
\label{eq:37}
 \ic{\Gamma}_{ij}\,\delta\lambda^{i}\,\delta\lambda^{j} \le n_{\Band}^{2}\ ,
\end{equation}
where $\delta\lambda^{i} = \ic{\lambda}^{i}_{\MLE} - \lambda^{i}_{s}$.
Note that the Fisher ellipse actually describes the fluctuations of
the maximum-likelihood estimator $\ic{\lambda}_{\MLE}$ for given
signal location. However, provided the likelihood-manifold is not
strongly curved, this also describes our uncertainty of the signal
location for given MLE $\ic{\lambda}_{\MLE}$, as indicated in
Fig.~\ref{fig:searchbox}.
The zoom stage will yield the fully-coherent maximum-likelihood
estimator $\co{\lambda}_{\MLE}$, which represents our best estimate
for the signal parameters $\lambda_{s}$.
Thus the two-stage search strategy corresponds to the transition
\begin{center}
 \begin{tikzpicture}[scale=0.75]

\draw -- (0,0) node {$\ic{\lambda}_{c}$};
\draw[->,black](0.3,0)--(2.9,0);
\draw -- (1.5,0.4) node {refinement};
\draw -- (3.5,0) node {$\ic{\lambda}_{\mathrm{MLE}}$};
\draw[->,black](4.1,0)--(6.8,0);
\draw -- (8.0,0) node {$\co{\lambda}_{\mathrm{MLE}}\approx\lambda_{s}\,.$};
\draw -- (5.3,0.35) node {zoom};

\end{tikzpicture}

\end{center}
In the following we use a subscript $R$ to denote quantities in the
refinement stage and a subscript $Z$ for quantities in the zoom stage.

The search volume for the refinement stage depends on the template
bank construction of the original semi-coherent search.
Ideally one iso-mismatch ellipse corresponding to the original
template-bank construction (see Fig.~\ref{fig:searchgrid}) should be
sufficient. In practice, however, it might often be neccessary to use
several grid spacings in each direction, if the template bank was not
originally constructed in a strictly metric way. In this case the
exact number of grid spacings will have to be empirically determined
in a Monte Carlo study.

\subsubsection*{Bounding box and volume of $n$-dimensional ellipses}
\label{sec:useful-relations-iso}

In the following it will be useful to express the bounding box
and volume of an $n$-dimensional ellipse, namely for the
iso-mismatch ellipse of Eq.~\eqref{eq:36} and the Fisher ellipse of Eq.~\eqref{eq:37}.
The general form of the $n$-dimensional ellipse equation is
\begin{equation}
  \label{eq:4}
  G_{ij} \,d\lambda^i\,d\lambda^j = R^2\,,
\end{equation}
where $G_{ij}$ is a positive-definite symmetric matrix.
The extents $\Delta\lambda^i$ of a bounding box along
coordinate axes $\lambda^i$ (as indicated in Fig.~\ref{fig:searchbox})
can be obtained from the diagonal elements of the inverse matrix,
$\{G^{-1}\}^{ij}$, namely
\begin{equation}
  \label{eq:5}
  \Delta\lambda^i = 2 R\,\sqrt{\{G^{-1}\}^{ii}}\,.
\end{equation}
The ellipse coordinate volume is expressible via the
matrix determinant, $\det G$, namely
\begin{equation}
  \label{eq:6}
  V = \frac{ R^n } { \sqrt{\det G} } \mV_n\,,
\end{equation}
where $\mV_n = \frac{\pi^{n/2}}{\Gamma(1 + n/2)}$ is the volume of unit $n$-ball.

\subsection{Classification of zoom outcomes}
\label{sec:class-zoom-outc}

Assuming a real CW signal, we can estimate the range of expected
values of the fully-coherent zoom $\mF$-statistic in $\co{\lambda}_{s}$. From Eq.~(\ref{eq:14}) we
can obtain a (rough) estimate of the average-$\SNRSQ$ from
the \emph{measured} average-$\SNRSQ$ of the semi-coherent
maximum-likelihood estimator, namely
\begin{equation}
\label{eq:40a}
 \avg{\snr^{2}}_{\MLE} \approx \avg{2\mF}_{\MLE} - 4\ .
\end{equation}
The $\SNRSQ$ of the fully-coherent search is linear in the number of
segments $\Nseg$, i.e.\
\begin{equation}
 \label{eq:40b}
 \co{\snr}^{2} = \Nseg\avg{\snr^{2}}_{\MLE}\ .
\end{equation}
Substitution of the above expression in Eq.~(\ref{eq:11}) yields
the expectation for the fully-coherent matched filter in
$\co{\lambda}_{\MLE}$, namely
\begin{equation}
 \label{eq:41a}
\co{2\mF}_o \equiv E[\co{2\mF}] \approx 4 + \Nseg\,\avg{\snr^{2}}_{\MLE}\ .
\end{equation}
Further substitution of Eq.~(\ref{eq:40b}) in Eq.~(\ref{eq:12}) yields
the corresponding variance as
\begin{equation}
 \label{eq:42a}
 \sigma^2_o \equiv \sigma^2[\co{2\mF}] \approx 2(4+2\Nseg\,\avg{\snr^{2}}_{\MLE})\ .
\end{equation}
These quantities are useful for defining what we mean by confirming a
CW signal.

Note that the uncertainty in the original SNR-estimation in
Eq.~\eqref{eq:40a} results in a distribution around the final estimate
of Eq.~\eqref{eq:41a} that is wider than estimated by Eq.~\eqref{eq:42a}.
This effect can be computed analytically and empirically, and is found
to amount to about a factor of 2.

Depending on the maximal $\co{2\mF}$ value found in the final zoom stage, we
can distinguish 3 possible outcomes:
\begin{itemize}

 \item \textit{Consistency with Gaussian noise ($\CGN$)} - the fully-coherent $\co{2\mF}$
value does not exceed a threshold
\begin{equation}
\label{eq:44a}
 \co{2\mF}<\CGNth\,,
\end{equation}
where $\CGNth$ is chosen to corresponds to some (small)
false-alarm probability $\pfA$ in Gaussian noise.

For example, a threshold $\CGNth=60$ corresponds to a
very small false-alarm probability of order $10^{-12}$ in a single
template, as given by Eq.~\eqref{eq:40}.

\item \textit{Non-Gaussian origin ($\NGO$)} - the candidate is loud enough to be
inconsistent with Gaussian noise at the false-alarm probability $\pfA$, i.e.
\begin{equation}
 \label{eq:44b}
\co{2\mF}\ge\CGNth\ .
\end{equation}

\item We define \emph{signal recovery} ($\SMC$) as a \emph{subclass} of $\NGO$,
namely if the final zoomed candidate $\co{2\mF}$
exceeds the Gaussian-noise threshold $\CGNth$
\emph{and} falls into the predicted signal interval given by Eqs.~\eqref{eq:41a} and
\eqref{eq:42a} (at some confidence level).
We can write this as
\begin{equation}
 \label{eq:43a}
  \coFth < \co{2\mF} < \coFmax
\end{equation}
where $\coFth \equiv \max\{ \CGNth, \, \co{2\mF}_o -
n_{u}\,\sigma_o\}$, and
$\coFmax\equiv \co{2\mF}_o + n_{u}\,\sigma_o$, where
$n_{u}$ determines the desired confidence level. In this work we
consider $n_{u}=6$, which corresponds roughly to a confidence of$~\sim99.6\%$.

\end{itemize}

Note that there can be cases where a zoomed candidate ends up in $\NGO$
but does not make it into the signal recovery ($\SMC$) band,
e.g. typically
$\CGNth < \co{2\mF} < \coFth$.
There can be different reasons for this, e.g. the search algorithm converged to a
secondary maximum in the refinement or zoom stage, the signal model
deviates from reality and requires modification, or  the ``signal''
found is of non astrophysical origin (eg a detector-noise artifact).
Generally further investigation will be required for all candidates
falling into the non-Gaussian category ($\NGO$).

\subsection{Grid-based computing-cost of the zoom stage}
\label{sec:comp-cost-estim-grid-based}

We do not consider a grid-based follow-up method in this paper, but it is
instructive to estimate the corresponding computing-cost for later comparison.
To estimate the number of templates required for the fully-coherent
search, we can use Eq.~\eqref{eq:6} to compute the volume of the
follow-up Fisher ellipse, Eq.~\eqref{eq:37}, and divide it by the volume
covered by one coherent template, Eq.~\eqref{eq:36}.
Namely, the Fisher-ellipse volume is given by
\begin{equation}
  \label{eq:2}
  \ic{V} = \frac{n_{\Band}^n}{\big(\Nseg\,\avg{\rho^{2}}\,\big)^{n/2}\,\sqrt{\det \ic{g}}}\,\mV_n\,,
\end{equation}
while the coherent template-volume at mismatch $m$ is
\begin{equation}
  \label{eq:7}
  \co{V} = \frac{m^{n/2}}{\sqrt{\det \co{g}}}\,\mV_n\,,
\end{equation}
therefore we can estimate then number of template as
\begin{equation}
  \label{eq:8}
  \Ntemp \approx \frac{\ic{V}}{\co{V}} =
  \frac{n_{\Band}^n}{\big(\Nseg\,\avg{\rho^2}\,\big)^{n/2}\,m^{n/2}}{\frac{\sqrt{\det \co{g}}}{\sqrt{\det\ic{g}}}}\,.
\end{equation}
Consider a follow-up of a candidate from a directed $n=2$ search in
$\{f,\dot{f}\}$ (e.g. see Fig.~\ref{fig:searchgrid}).
Assuming a semi-coherent search using
$\Nseg = 200$ segments of $\Tseg = 1\ \days$ duration without gaps,
and a fully-coherent observation time of $T=200\ \days$.
Using the expressions found in \cite{Pletsch:2010a}, the determinants
of the two-dimensional coherent and the semi-coherent metrics are
found as
 \begin{align}
  \label{eq:34b}
  \sqrt{\det\co{g}} &= \pi^{2}T^{3} \,\frac{1}{540}\,,\\
  \sqrt{\det\ic{g}} &= \pi^{2}\Tseg^{3} \,\frac{\gamma(\Nseg)}{540}\,,
\end{align}
where $\gamma\approx \sqrt{5}\,\Nseg$ is the spindown refinement factor.
Putting everything together in Eq.~\eqref{eq:8}, we obtain
\begin{equation}
\label{eq:35b}
 \Ntemp \approx \frac{n_{\Band}^{2}\Nseg}{\sqrt{5}\,\avg{\rho^2}\, m}\ .
\end{equation}
where we used $\Nseg=T/\Tseg$.
For a signal with $\avg{\rho^{2}}= 1$, $n_{\Band}=24$
\footnote{This large $n_{\Band}$ value is found to contain the signal location
in more than 98 \% of the cases even for weak signals, where the Fisher-matrix 
may be a poor predictor, see \cite{PhysRevD.77.042001}.}
 and $m=0.1$ the
number of templates is therefore $\Ntemp\approx5.1\times10^{5}$.
Thus, using Eq.~(\ref{eq:30}) for 2 detectors and the SFT method with $\Tsft = 1800\ \seconds$,
the zoom computing cost is $\CC\approx11\ \minutes$ per
candidate, where we used the
fundamental computing cost $\cOSFT=7\times10^{-8}\ s$\cite{PrixShaltev2011}.

In the more general case where the sky position of the source is also unknown,
the number of sky points typically scales at least quadratically with
the observation time  \cite{PhysRevD.79.022001,Pletsch:2010a} (for
coherent integration longer than few days), thus generally resulting
in completely prohibitive  computational requirements for grid-based
follow-up searches. In particular, extending the directed search example
from the previous paragraph to an all-sky follow-up would require
 $\Ntemp_{\mathrm{sky}}\approx1.3\times10^{6}$ sky points\footnote{The number of sky templates has been estimated by
 numerical computation of the sky part of the metrics $\ic{g}$ and $\co{g}$
 using \texttt{FstatMetric\_v2} from LALSUITE \cite{LALSuite:Misc}, see also
 \cite{prix06:_searc}.}, or a total of $\Ntemp\approx6.8\times10^{11}$
 templates.

For comparison, using the grid-less search algorithm discussed in the
next sections, it is possible to coherently follow up 2-D directed
candidates in less than 2 minutes, see Fig.~\ref{fig:dms_200_d},
and all-sky candidates in about $1$~hour per candidate, see
Fig.~\ref{fig:ams_200_d}.

\subsection{Mesh Adaptive Direct Search (MADS)}
\label{sec:mesh-adaptive-direct}

A significant difference between the hierarchical search strategies discussed
in \cite{Brady:1997ji,Brady:1998nj,PhysRevD.72.042004} and this work is the method
of template bank construction at the different stages. Namely, we
consider a grid-less method for exploring the parameter space.

The MADS class of algorithms for derivative-free optimization has been
first introduced in \cite{Audet04meshadaptive} and further developed in \cite{AbramsonADD09}
and \cite{LeDigabel2011} among others. In this subsection we only introduce some of the
control parameters of the algorithm required in the construction of
MADS-based
$\mF$-statistic searches, for an in-depth treatment and proofs we refer
the reader to the cited publications.

MADS consist of the iteration of two steps, called \textit{search} and \textit{poll},
in which trial points
are constructed and evaluated in order to find an extremum. In the search step
any strategy can be applied to construct trial points. In this work we use
quadratic models (quadratic form) to approximate the objective function from a sample
of objective values \cite{LeDigabel2011}. If the local exploration
in the search step fails to generate a new solution, a set of poll
points is generated using a stochastic or deterministic
method. Stochastic  means that the poll points are generated randomly
 \cite{Audet04meshadaptive}, where deterministic refers  to the usage of
 pseudo-random Halton sequences \cite{AbramsonADD09}.
However both methods generate points which form a
dense set in the unit sphere after an infinite number of iterations.
 For a given starting point $\lambda_{c}$ with parameter space boundaries
$\Delta\lambda_{\Band}$, initial step sizes $d\lambda$ and a method
for generation of poll points,
the discretization of the parameter space $\Delta^{m}_{k}$ at iteration $k$ is governed
by a fixed rational number $\mub>1$ and the coarsening $\mce\ge0$ and
refining $\mre\le-1$ exponents. If the current iteration generates a better
solution, the discretization in the next iteration is coarser, namely
 $\Delta^{m}_{k+1}=\mub^{\mce}\Delta^{m}_{k}$, otherwise
$\Delta^{m}_{k+1}=\mub^{\mre}\Delta^{m}_{k}$ \cite{Audet04meshadaptive}. The
algorithm stops if an improved solution cannot be found or the total number of
evaluated parameter space points $\maxbbeval$ reaches some given maximum $\maxbbeval_{\max}$.

\subsection{MADS-based follow-up algorithm}
\label{sec:mads-based-search}

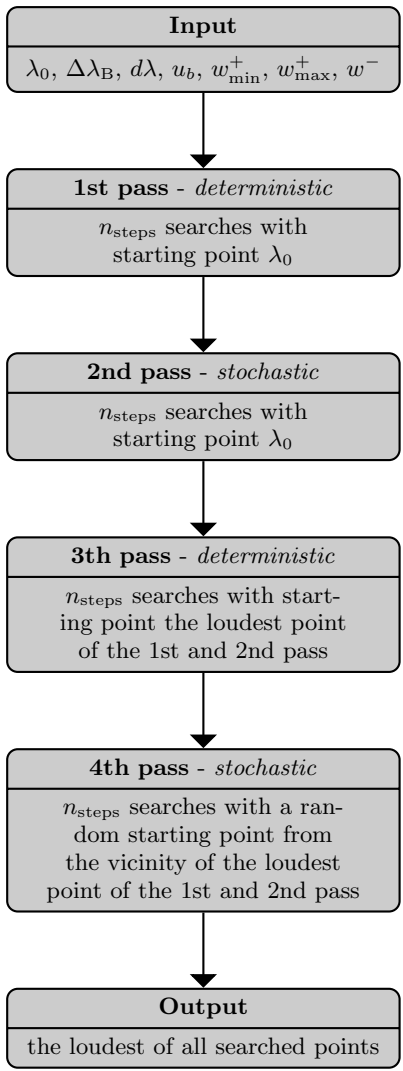
\begin{figure}
\tikzstyle{block}=[rectangle, draw=black, rounded corners, fill=gray!40, text centered, anchor=north, text=black, text width=5cm,thick]
\tikzstyle{farrow}=[->, >=triangle 90, thick]
\begin{tikzpicture}[scale=0.75]

\node (Input) [block, rectangle split, rectangle split parts=2]
        {
            \textbf{Input}
	    \nodepart{second}$\lambda_{0}$, $\Delta\lambda_{\mathrm{B}}$, $d\lambda$, $u_{b}$, $w^{+}_{\mathrm{min}}$, $w^{+}_{\mathrm{max}}$,  $w^{-}$
        };

\node (Pass1) [block, rectangle split, rectangle split parts=2, below=of Input]
        {
            \textbf{1st pass} - \textit{deterministic}
            \nodepart{second}$n_{\mathrm{steps}}$ searches with starting point $\lambda_{0}$
        };

\node (Pass2) [block, rectangle split, rectangle split parts=2, below=of Pass1]
        {
            \textbf{2nd pass} - \textit{stochastic}
            \nodepart{second}$n_{\mathrm{steps}}$ searches with starting point $\lambda_{0}$
        };

\node (Pass3) [block, rectangle split, rectangle split parts=2, below=of Pass2]
        {
            \textbf{3th pass} - \textit{deterministic}
            \nodepart{second}$n_{\mathrm{steps}}$ searches with starting point the loudest point of the 1st and 2nd pass
        };

\node (Pass4) [block, rectangle split, rectangle split parts=2, below=of Pass3]
        {
            \textbf{4th pass} - \textit{stochastic}
            \nodepart{second}$n_{\mathrm{steps}}$ searches with a random starting point from the vicinity of the loudest point of the 1st and 2nd pass
        };

\node (Output) [block, rectangle split, rectangle split parts=2, below=of Pass4]
        {
            \textbf{Output}
            \nodepart{second}the loudest of all searched points
        };

\draw[farrow] (Input.south) -- ++(0,0) -| (Pass1.north);
\draw[farrow] (Pass1.south) -- ++(0,0) -| (Pass2.north);
\draw[farrow] (Pass2.south) -- ++(0,0) -| (Pass3.north);
\draw[farrow] (Pass3.south) -- ++(0,0) -| (Pass4.north);
\draw[farrow] (Pass4.south) -- ++(0,0) -| (Output.north);

\end{tikzpicture}
\caption{MADS-based search algorithm with 4 passes, where $\lambda_{0}=\lambda_{c}$ in the refinement stage and $\lambda_{0}=\ic{\lambda}_{\MLE}$ in the zoom stage.}
\label{fig:alg}
\end{figure}

From the point of view of the MADS algorithm, the function to optimize is a black-box
requiring some input to produce a single output value. The black-box
in our case is either the computation of the semi-coherent
$\mF$-statistic $\avg{2\mF}$ of Eq.~(\ref{eq:13}) in the refinement,
or the fully-coherent $\mF$-statistic $\co{2\mF}$ in the zoom stage.
In order to minimize the possibility of convergence to secondary maxima, we run multiple instances
of the MADS search in each stage varying the mesh coarsening exponent $\mce$. The minimal $\minmce$ and maximal $\maxmce$ coarsening
 exponent  determine the number
of MADS steps in each pass, namely $\nsteps = \maxmce-\minmce+1$.
Thus we consider our search algorithm
composed of several instances of MADS, see Fig.~\ref{fig:alg}.
The input of the search algorithm is the candidate $\lambda_{c}$ to follow-up,
the search boundaries $\Delta\lambda_{R/Z}$  around the candidate
and a set of MADS input parameters, namely
$\left\{d\lambda,\mub,\minmce,\maxmce,\mre\right\}$.
In the zoom stage the search boundaries ($\Delta\lambda_{Z}$) are
estimated from the bounding box of the Fisher ellipse, using Eq.~\eqref{eq:5}.
For the refinement stage the search boundaries ($\Delta\lambda_{R}$)
generally have to be determined depending on the template-bank setup
of the original semi-coherent search.
Note, however, that the bounding boxes $\Delta\lambda$ only serve
as a necessary input parameter to the NOMAD search algorithm, while
the effective MADS search region can be further reduced by
rejecting points that do not satisfy a given constraint.
For example, the effective search region in the zoom stage always
consists of the Fisher ellipse Eq.~\eqref{eq:37}.

The initial step sizes $d\lambda^i$ are also empirically determined,
typically as some fraction of the search boundary $\Delta\lambda_{R/Z}^{i}$.

We propose a 4 pass algorithm with equal (for simplicity) number of steps $\nsteps$
in each pass, however with different starting point and method of trial-point generation:
\begin{itemize}
 \item \textit{1st pass} - starting point $\lambda_{c}$, \deterministic
point generation,
 \item \textit{2nd pass} - starting point $\lambda_{c}$, \stochastic
point generation,
 \item \textit{3rd pass} - starting point loudest template from
the first 2 passes, \deterministic point generation,
 \item \textit{4th pass} - starting point from the vicinity
of the loudest point from the first 2 passes, \stochastic point generation.
\end{itemize}
In the zoom stage we terminate the search as soon as the loudest point
of the current iteration satisfies the signal-confirmation condition
($\SMC$) of Eq.~\eqref{eq:43a}. In lower-dimensional cases, such as the
directed search considered later, a single pass is therefore often
found to be sufficient.
For later usage we introduce the total number of MADS iterations
$n_{I}$ as the sum of the number of steps in each pass.

\subsection{MADS-followup computing cost}
Contrary to grid-based searches, the computing cost of the MADS based
algorithm is non-deterministic, due to the a-priori unknown number of
explored parameter space points.
To estimate the maximal computing cost of the refinement or the zoom stage using
Eq.~(\ref{eq:27}), we need the maximal number of possibly evaluated templates
\begin{equation}
\label{eq:42}
 \Ntemp_{\mathrm{max}} = \sum_{i=0}^{n_{I}}\maxbbeval^{i}_{\max}\ ,
\end{equation}
where $\maxbbeval^{i}_{\max}$ is the user-specified maximum of the number of
computed templates at MADS iteration $i$.
This maximal number is typically chosen large to avoid too early
interruption of the MADS instance, e.g. when further improvement of the current
solution is possible while the extremum is not yet found. However, if the extremum
is found, a MADS iteration starting from this point  terminates rapidly. 

Note that the fundamental computing cost $\cOSFT$ in stochastic
searches over the sky is typically larger than in a grid-based search,
where a lot of templates with different spindown components can be
computed at fixed sky position. This results in a larger value of about
$\cOSFT\approx3\times10^{-7}\ s$ instead of the number quoted in
Sec.~\ref{sec:comp-cost-estim-grid-based}.

\subsection{False-alarm and detection probability}

After the final fully-coherent zoom stage we are left with a candidate falling
in one of the three categories discussed earlier, namely: the candidate is
consistent with the signal model $(\SMC)$, with Gaussian noise $(\CGN)$ or
is of non-Gaussian origin but inconsistent with the signal model.
An additional valuable piece of information is the false-alarm probability associated with
the candidate. This is the  probability of exceeding a threshold $2\mF$-value in the
absence of a signal, where the relevant distribution is the central
 $\chi^{2}$ distribution with 4 degrees of freedom, denoted as $\chi^{2}_{4}(2\mF)$.
The single-template false-alarm probability is
\begin{eqnarray}
\label{eq:40}
 \pfA^{1}&=&\int_{2\Fth}^{\infty}d(2\mF)\chi^{2}_{4}(2\mF)\\\nonumber
 &=&(1+\Fth)e^{-\Fth}\ ,
\end{eqnarray}
and for $\Ntemp$ independent templates this results in
\begin{equation}
\label{eq:41}
 \pfA = 1 - \left(1-\pfA^{1}\right)^{\Ntemp}\ ,
\end{equation}
where for $\Ntemp\pfA^{1}\ll1$, Taylor expansion yields
$\pfA\approx\Ntemp\pfA^{1}$. For example,
a threshold of  $\CGNth = 70$ for a search with
$\Ntemp = 1\times10^{5}$ templates corresponds to a
false-alarm probability of
 $\pfA\lesssim2\times10^{-9}$, where the upper bound corresponds to $\Ntemp$
completely independent templates.

The overall detection probability of the follow-up method
depends on the signal SNR. Higher SNR in the refinement stage
 yields better localization of the signal, i.e. a smaller Fisher
 ellipse and thus also a higher probability of signal recovery (\ref{eq:43a}). In addition,
the MADS-algorithm parameters also affect the detection efficiency, e.g. an increased
number of MADS iterations increases the detection probability, especially for signals
with lower SNR. Because of this, the detection probability will have
to be estimated empirically in a Monte Carlo study, see
Figs.~\ref{fig:dms_200_c} and \ref{fig:ams_200_c}.

\section{Monte Carlo studies}
\label{sec:monte-carlo-study}
To demonstrate the capability of the systematic follow-up procedure
proposed in Section \ref{sec:madsforcgw} we perform two different
types of Monte Carlo (MC) studies.

In the first case we simulate a so-called \emph{directed} search for
a fixed sky position, where we follow up candidates in a 2-dimensional
spindown space, i.e. $\{f,\dot{f}\}$.
In the second case we simulate an all-sky search over the
4-dimensional parameter space $\{\alpha, \delta, f, \dot{f}\}$.

All MADS searches are implemented using the MADS reference
library NOMAD \cite{LeDigabel2011A909} and the LAL library from the LALSUITE
\cite{LALSuite:Misc} is used for the $\mF$-statistic computation \cite{prix:_cfsv2}. The Gaussian
data and signal injections are produced using the LALAPPS programs from LALSUITE. In particular with \texttt{lalapps\_Makefakedata\_v4}
we create data sets of total duration $T = 200\ \days$, with $\Nseg=200$
segments of duration $\Tseg=1\ \days$, using SFTs of length
$\Tsft = 1800\ \mathrm{s}$, for the two LIGO detectors H1 and L1. The
noise level per detector is generated as Gaussian white noise with a
power-spectral density $\detnoise$ of $\sqrt{\detnoise}=2\times10^{-23}\,\Hz^{-1/2}$.

\begin{figure*}[htp]
  \subfloat[\;SNR loss of the initial candidates $\mm$ versus semi-coherent
metric mismatch $\mty$.]{\includegraphics[width=0.96\columnwidth]{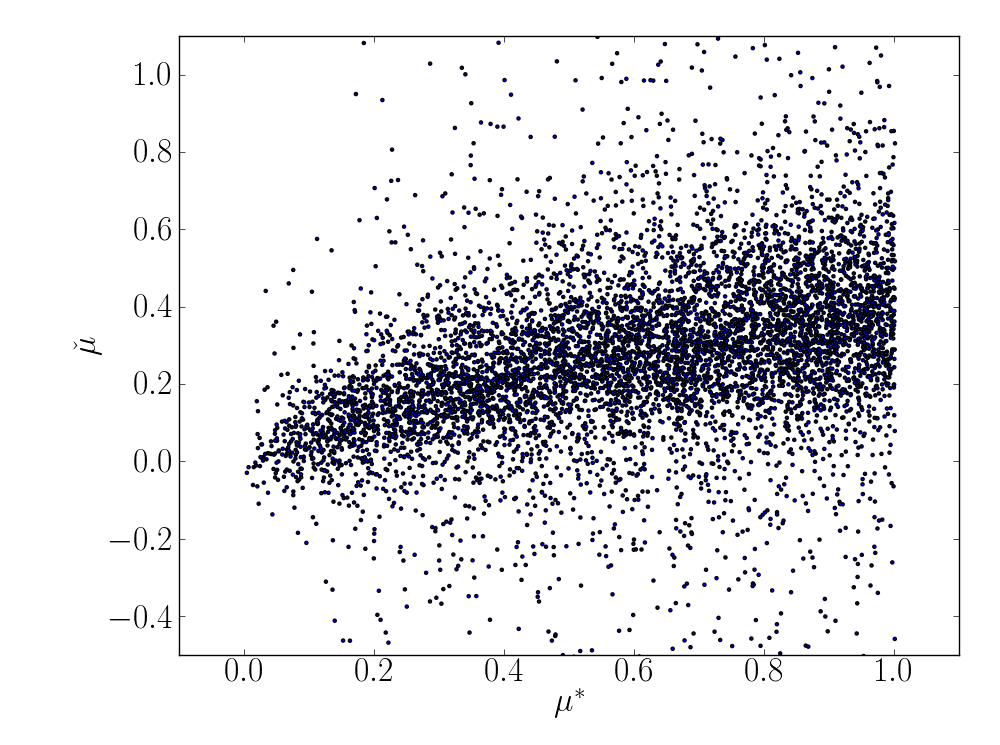}\label{fig:ams_200_a}}
\quad
\subfloat[$\co{2\mF}_{Z}$ distribution after the fully-coherent 4-D $\{\alpha,\delta,f,\dot{f}\}$ zoom stage of
\ASNT searches in pure Gaussian noise, without injected signal. The maximal
$2\mF$-value found is $\co{2\mF}_{Z}^{max}=\protect\input{AMS_200_noise_dist_max.tex}$.
 The mean value is $\tavg{\co{2\mF}_{Z}}=\protect\input{AMS_200_noise_dist_avg.tex}$ indicated with dots.]
{\includegraphics[width=0.9\columnwidth]{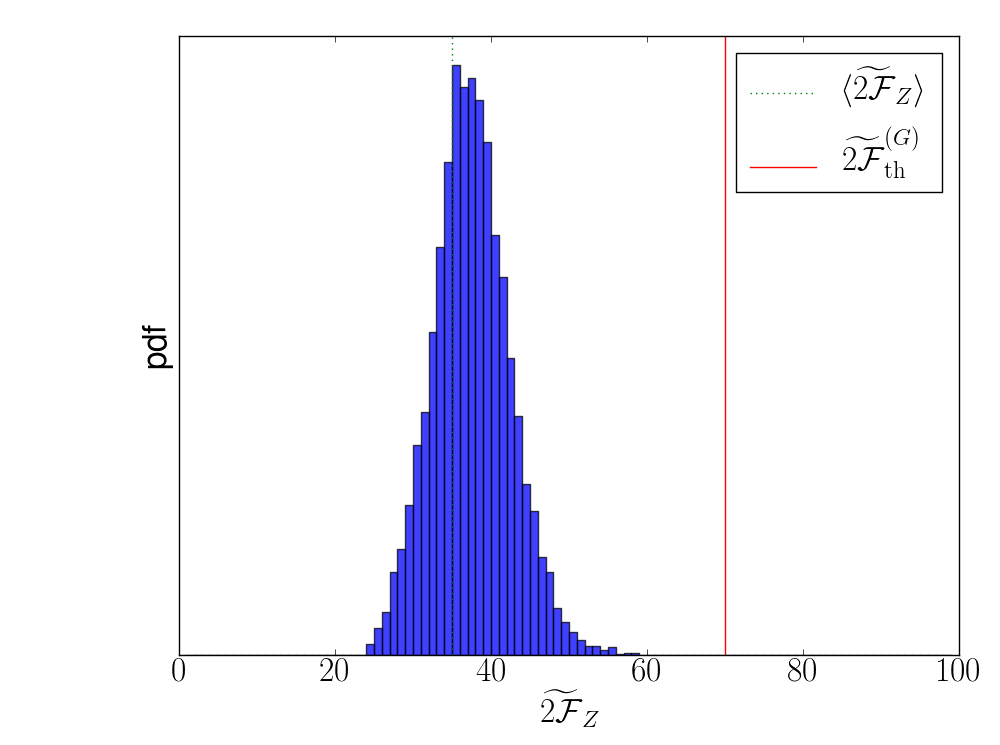}\label{fig:ams_200_b}}
\quad
\subfloat[\;Percentage of the \ASNT injected signals classified as recovered
(\textcolor{red}{$-$} $\ \SMC$)
and of non-Gaussian origin ($\textcolor{green}{\times}\ \NGO$)
as function of the signal strength $\avg{\snrsq_{s}}$. The error bars are
computed using a Jackknife estimator.]
{\includegraphics[width=0.96\columnwidth]{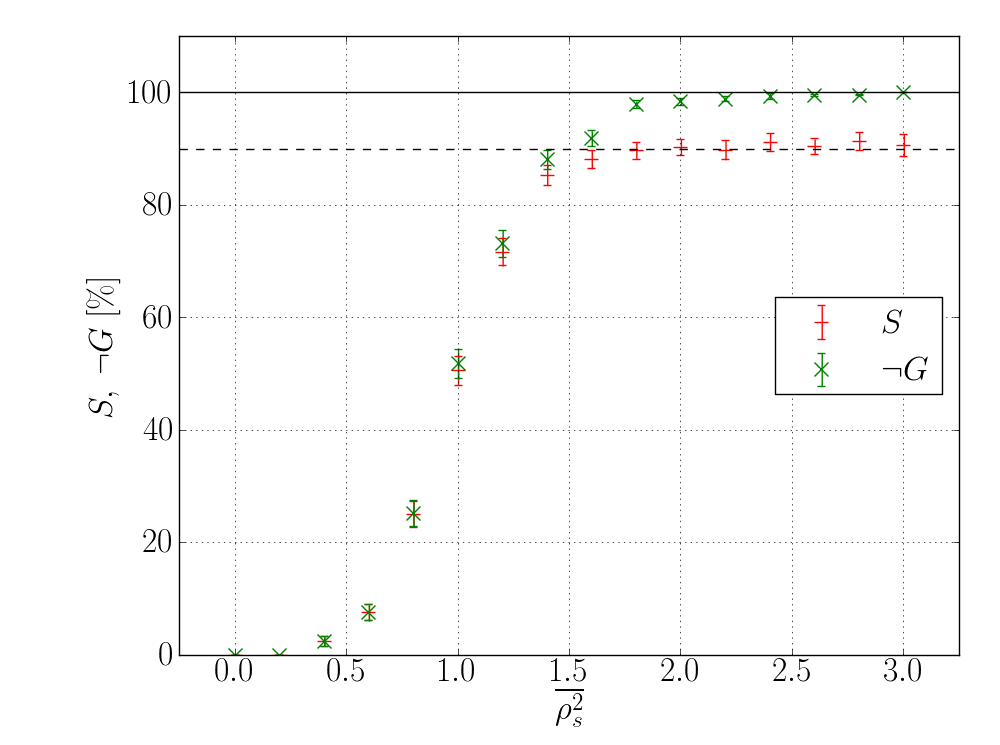}\label{fig:ams_200_c}}
\quad
\subfloat[\;\textit{Upper plot}: computing cost of the semi-coherent refinement stage.
 \textit{Middle plot}: computing cost of the fully-coherent zoom stage.
\textit{Lower plot}: total computing cost.
]{\includegraphics[width=0.96\columnwidth]{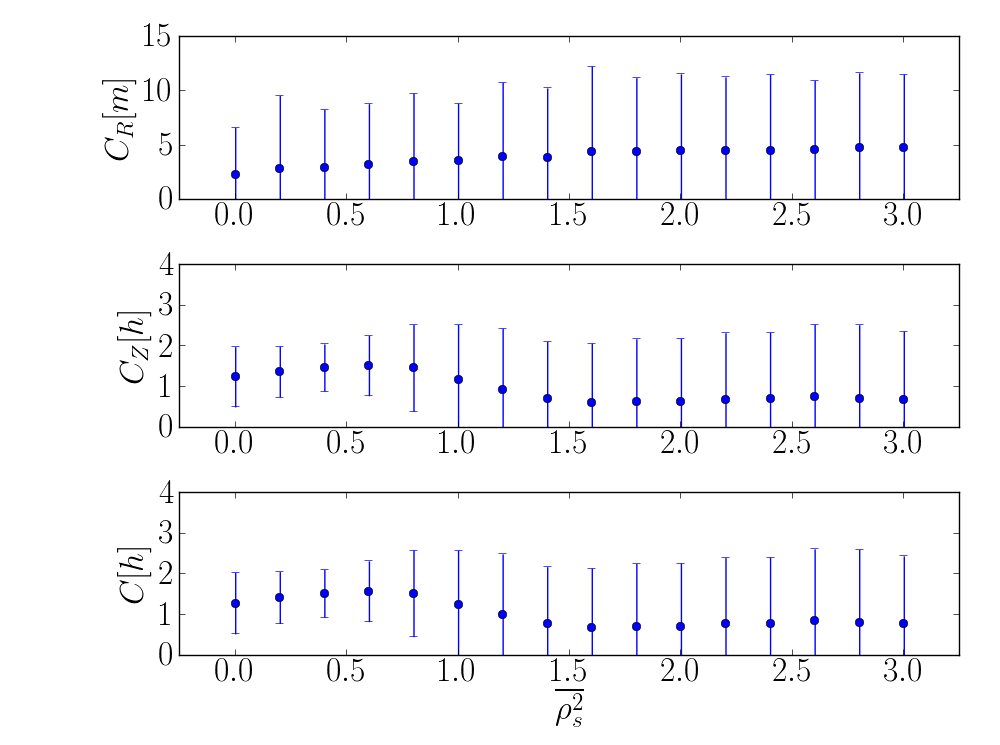}\label{fig:ams_200_d}}
\caption{Monte Carlo study of 2-stage follow-up of candidates from an
  all-sky $\{\alpha,\delta,f,\dot{f}\}$
semi-coherent search with $N=200$ segments of duration $\Tseg=1 \days$.}
\label{ignore-fig:ams_200}
\end{figure*}

Independently of the type of the search, the initial candidates to follow-up are
prepared as follows. Rather than performing a semi-coherent grid based search
using the Hough- \cite{Krishnan:2004sv} or GCT-method \cite{2009PhRvL.103r1102P},
we generate candidates by drawing a random point in the vicinity of the injection and
consider it a candidate if the semi-coherent metric mismatch $\mty$ is within the range
\begin{equation}
 \label{eq:mty}
\mty\in[0,1]\ ,
\end{equation}
see Figs. ~\ref{fig:dms_200_a} and ~\ref{fig:ams_200_a}.
This procedure for candidate
preparation allows us to separate the study of the follow-up algorithm from the
problem of how to setup a semi-coherent search, which is a difficult question on its own.

Note that even if the original grid-based semi-coherent search does not produce
candidates that conform with Eq.~(\ref{eq:mty}), we can always increase the density
of the grid until (\ref{eq:mty}) applies. This would
amount to a (cheap) pre-processing stage inserted before the present follow-up procedure .

\subsection{Follow-up of candidates from a directed search}
\begin{figure*}[htp]
\subfloat[\;SNR loss of the initial candidates $\mm$ versus semi-coherent
metric mismatch $\mty$ to the closest template.]
{\includegraphics[width=0.96\columnwidth]{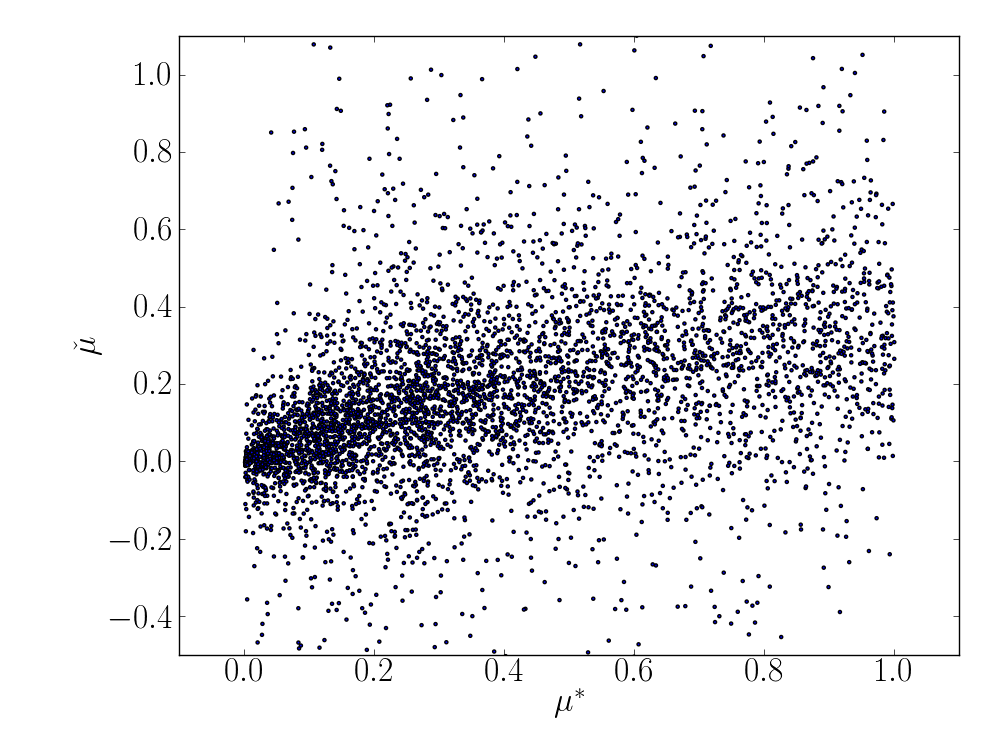}\label{fig:dms_200_a}}
\quad
\subfloat[$\co{2\mF}_{Z}$ distribution after the fully-coherent 2-D $\{f,\dot{f}\}$
zoom stage of  \DSNT directed searches  in pure Gaussian noise without injected signal. The maximal $2\mF$-value found is $\co{2\mF}_{Z}^{max}=\protect\input{DMS_200_noise_dist_max.tex}$. The mean value $\tavg{\co{2\mF}_{Z}}=\protect\input{DMS_200_noise_dist_avg.tex}$ is plotted with a dotted line.]
{\includegraphics[width=0.96\columnwidth]{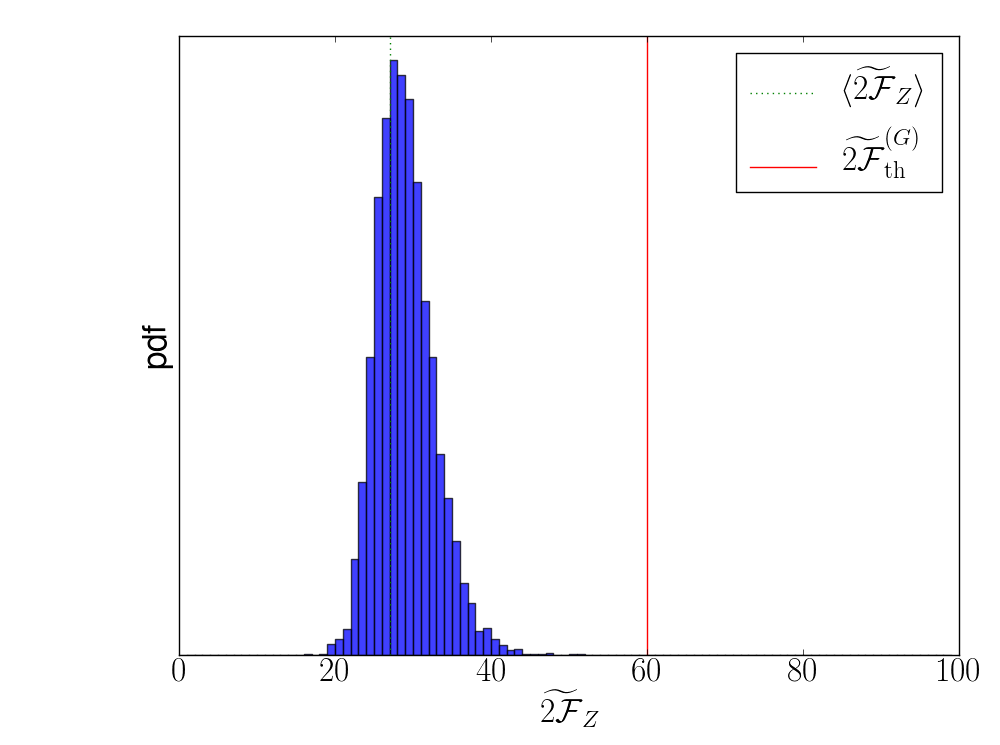}\label{fig:dms_200_b}}
\quad
\subfloat[\;Percentage of the \DSNT injected signals classified as recovered
(\textcolor{red}{$-$} $\ \SMC$)
and of non-Gaussian origin ($\textcolor{green}{\times}\ \NGO$) as
function of the non-centrality parameter  $\avg{\snrsq_{s}}$,
 Eq.~(\ref{eq:16}).
 The error bars are computed by using a Jackknife estimator.]
{\includegraphics[width=0.96\columnwidth]{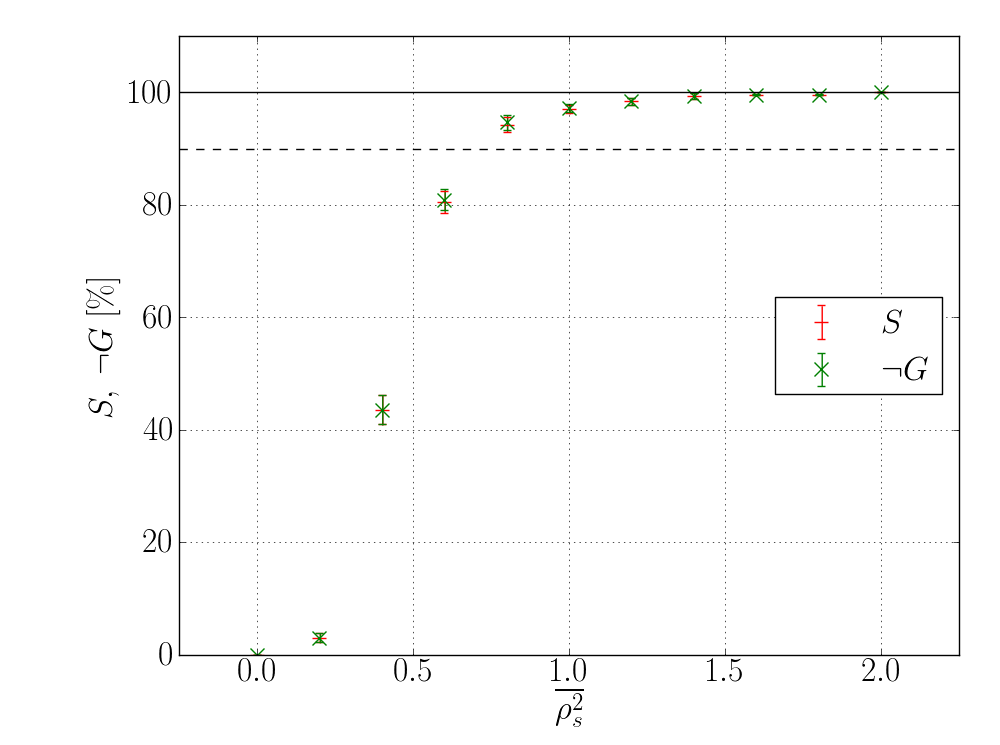}\label{fig:dms_200_c}}
\quad
\subfloat[\;\textit{Upper plot}: computing cost of the semi-coherent refinement stage.
 \textit{Middle plot}: computing cost of the fully-coherent zoom stage.
\textit{Lower plot}: total computing cost.
]
{\includegraphics[width=0.96\columnwidth]{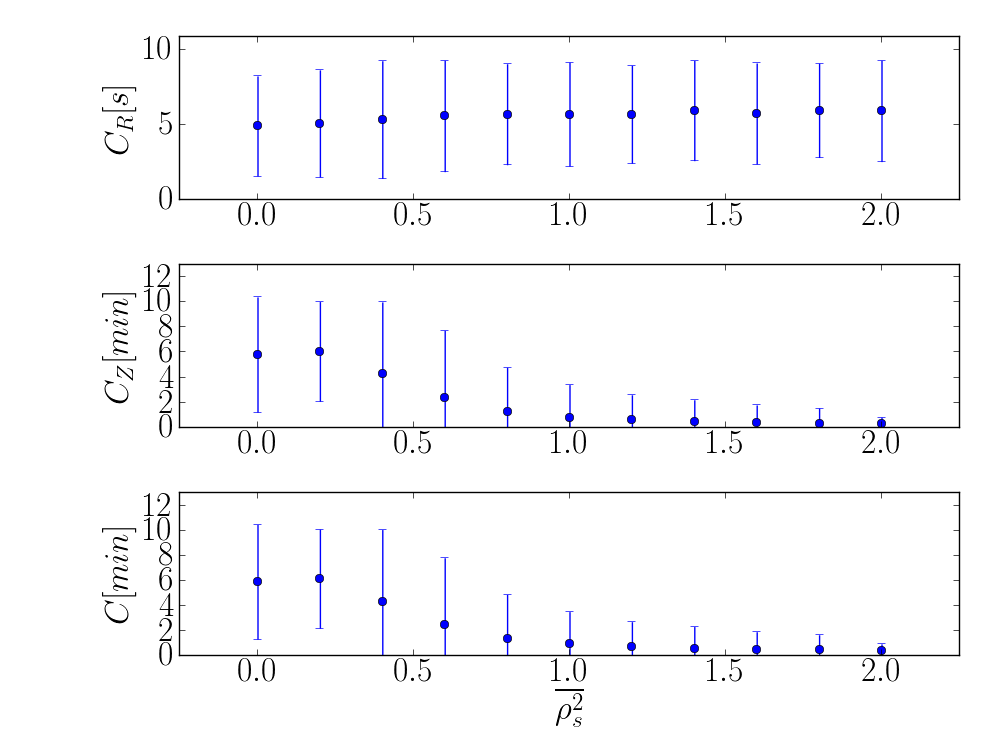}\label{fig:dms_200_d}}
\caption{Monte Carlo study of 2-stage follow-up of candidates from a directed
$\{f,\dot{f}\}$ semi-coherent search pointed toward the Galactic Center with $\Nseg=200$ segments
of duration $\Tseg=1\ \days$.}
\label{fig:dms_200}
\end{figure*}
For the directed type of searches we fix the sky position to the coordinates of
the Galactic Center. This choice is arbitrary and we could use any other
 point without qualitatively changing the results. We create
\DSNT data sets. Note that each data set has different Gaussian
noise realization in which a CW signal from an isolated source is injected. In the
process of injection, the original noise data set is also used to
examine the behavior of the follow-up method in the absence of a signal.

The pulsar injection parameters $\lambda_{s}$ are drawn uniformly in the range $f\in(50,51)\ \Hz$, $\cos\iota\in(-1,1)$,
 $\psi\in(-\pi/4,\pi/4)$ and $\phi_{0}\in(0,2\pi)$. The signal amplitude $h_{0}$
is chosen such that the expected average-$\SNRSQ$ of Eq.~\eqref{eq:16}
for a perfect match is distributed uniformly in the range $\avg{\snrsq}_{s}\in(0,2)$. The
spindown $\dot{f}$ is chosen uniformly in the range
$\dot{f}\in(-\frac{f_{\min}}{\tau_{\min}},\frac{f_{\min}}{\tau_{\min}})$ with
minimal spindown age $\tau_{\min}=300\ \years$ at $f_{\min}=50\ \Hz$.
The MADS-algorithm parameters used in the MC are summarized in Table
\ref{tab:mc-parameters}, which have been found empirically to
achieve good results.
\begin{table}[htbp]
 \centering
\begin{tabular}[c]{c|c|c|c|c|c}
stage & $\mre$ & $\minmce$ & $\maxmce$ & $\mub$ &
$\maxbbeval_{\mathrm{max}}$ \\\hline
$R$ & -1 & 1 & 1 &  2 & 20000  \\\hline
$Z$ & -1 & 1 & 50 &  1.1 & 20000
\end{tabular}
\caption{Algorithm parameters for follow-up of candidates from directed searches.}
\label{tab:mc-parameters}
\end{table}
For this type of follow-up we find that the 1st pass of the search algorithm in the
refinement stage and only two repetitions of the 2nd pass in the zoom stage is sufficient.
We restrict the size of the search box  for the refinement stage $\Delta\lambda_{R}$
by taking 1 frequency  and 2 first spindown  metric extents.
In the zoom stage we constrain the parameter space to a Fisher ellipse
Eq.~\eqref{eq:37} with  $n_{\Band}=24$.

We first apply the follow-up chain to the pure Gaussian-noise data without injected signals.
The corresponding $\co{2\mF}_{Z}$ distribution of the resulting fully-coherent
zoom stage is plotted in Fig.~\ref{fig:dms_200_b}. The maximal value found is
$\co{2\mF}_{Z}^{max}=\protect 51.61
$.
We therefore use a threshold for the classification of non-Gaussian
candidates $(\NGO)$ of $\CGNth=60$, which is safely above this level.

We next apply the follow-up chain to the Gaussian-noise data with injected signal.
In Fig.~\ref{fig:dms_200_c} we plot the percentage of injected signals
that are classified as recovered signals
$(\SMC)$ and non-Gaussian origin $(\NGO)$, as a function of the
injected signal strength $\avg{\snrsq}_{s}$. From this plot we can read out the detection probability,
namely we reach $90\%$ of signal recovery for candidates with $\avg{\snrsq_{s}}\approx0.7$.

The computing cost as function of $\avg{\snrsq_{s}}$ is plotted in Fig.~\ref{fig:dms_200_d}. We notice
that the cost of the refinement stage is negligible and in the zoom stage 
the averaged computing time decreases with higher signal strength.


\subsection{Follow-up of candidates from an all-sky search}
The data and signal preparation for the following all-sky Monte Carlo study is
the same as in the directed search case, however the sky position is
drawn isotropically over the whole sky. We create \ASNT data sets with uniformly
distributed injected average-$\SNRSQ$ in the range $\avg{\snrsq_{s}}\in(0,3)$.
The algorithm parameters used in the refinement and zoom stage are given in
Table \ref{tab:mc-parameters-allsky-2stage}, which have been
found empirically to yield good performance. We also find that here the zoom
stage benefits from performing all 4 search passes shown in
Fig.~\ref{fig:alg}.
The size of the search box for the refinement stage in the
spindown subspace has been defined exactly as in the directed
search example. The sky subspace is constrained
by using an $m=1$ iso-mismatch ellipse.
As in the previous example we use
$n_{\Band}=24$ in Eq.~\eqref{eq:37} to determine the size of the
Fisher ellipse.

\begin{table}[htbp]
 \centering
\begin{tabular}[c]{c|c|c|c|c|c}
stage & $\mre$ & $\minmce$ & $\maxmce$  & $\mub$ & $\maxbbeval$\\\hline
$R$ & -1 & 1 & 5 &  2 & 20000 \\\hline
$Z$ & -1 & 1 & 50 &  1.2 & 20000
\end{tabular}
\caption{Follow-up algorithm parameters for full parameter space searches.}
\label{tab:mc-parameters-allsky-2stage}
\end{table}

Similarly to the directed follow-up, we first test the pipeline using the Gaussian noise data without  injections.
The resulting distribution of final $\co{2\mF}_{Z}$ values is plotted in Fig.~\ref{fig:ams_200_b}.
The maximal value found is $\co{2\mF}^{max}_{Z}=\protect 58.76
$,
  which is higher compared to the
 value found  in the directed follow-up searches due to the increased
number of evaluated templates.
We therefore use a threshold for the classification of non-Gaussian
candidates $(\NGO)$ of $\CGNth=70$, which is safely above this level.


Next we search the data containing the injected signals.
In Fig.~\ref{fig:ams_200_c} we plot the fraction of
signals classified as recovered $(\SMC)$ and the percentage of MC trials
found to be of non-Gaussian origin $(\NGO)$, as a function of the
injected signal strength $\avg{\snrsq}_{s}$
In order to achieve $90 \%$ signal recovery ($\SMC$), we now need stronger signals, 
namely $\avg{\snrsq_{s}}\gtrsim1.7$.
However, for $\avg{\snrsq_{s}}\approx 1.5$ we can already achieve
$90\%$ ``detection probability'' in the sense of separating candidates
from Gaussian noise ($\NGO$).
This indicates that the zoom step sometimes
converges on a secondary maximum. Given that any non-Gaussian ($\NGO$)
candidates after zoom will receive further scrutiny, it would be
straightforward to further explore the parameter space around such
candidates to localize a potential primary maximum.

 The computing cost as a function of $\avg{\snrsq_{s}}$ is plotted in Fig.~\ref{fig:ams_200_d}. We notice
that the total computing cost is dominated by the zoom stage and the
averaged computing time is rather independent of the signal strength.


\section{Discussion}
\label{sec:disc}
We have studied a two-stage scheme for the fully-coherent follow-up of
semi-coherent candidates.
The first stage, called refinement, aims to find the maximum-likelihood
estimator of the initial semi-coherent candidate. This allows us to better constrain the
parameter space for the coherent zoom stage. The two-stage scheme is suitable
for following-up candidates from all-sky or directed semi-coherent searches. The proposed grid-less optimization
 lowers the computing cost per candidate to acceptable levels.
In Monte Carlo studies we tested the efficiency of the algorithm for directed and
all-sky follow-up searches.

In this paper we restricted the all-sky follow-up optimization to 4 dimensions,
namely sky, frequency and first spindown. Further work is required to
extend the optimization to higher dimensions. A related attractive
direction for further development is the extension and application of
the search algorithm for follow-up of CW candidates in binary systems,
which is a challenging higher-dimensional problem.

We also aim to extend the two-stage scheme presented here by including
intermediate semi-coherent zoom stages. This should allow
to further reduce the computing cost and increase detection efficiency.

\section{Acknowledgments}
We are grateful to numerous colleagues for useful comments and discussions, in particular
Karl Wette, Holger Pletsch, Paola Leaci, Vladimir Dergachev, Thomas Dent, Badri Krishnan, Maria Alessandra
Papa and Bruce Allen. MS gratefully acknowledges the support of Bruce Allen and
the IMPRS on Gravitational Wave Astronomy of the Max-Planck-Society. This paper
has been assigned LIGO document number \dcc.

\bibliography{ppfollowup}

\end{document}